# A Novel Approach of using AR and Smart Surgical Glasses Supported Trauma Care

Lal, Anurag [3], Ming-Hsien Hu[2], Pei-Yuan Lee[2], Wang, Min Liang[1,3*]

Department of Electrical Engineering, National Chung Cheng University, Taiwan(R.O.C)[1]

Department of Orthopedics, Show Chwan Memorial Hospital, Changhua 50544, Taiwan[2]

Taiwan Main Orthopaedic Biotechnology, Co. Ltd, Taichung City, 418, Taiwan[3]

*Authors to whom correspondence should be addressed.

Abstract

- BACKGROUND: Augmented reality (AR) is gaining popularity in varying field such as computer gaming and medical education fields. However, still few of applications in real surgeries. Orthopedic surgical applications are currently limited and underdeveloped.

- METHODS: The clinic validation was prepared with the currently available AR equipment and software. A total of 1 Vertebroplasty, 2 ORIF Pelvis fracture, 1 ORIF with PFN for Proximal Femoral Fracture, 1 CRIF for distal radius fracture and 2 ORIF for Tibia Fracture cases were performed with fluoroscopy combined with AR smart surgical glasses system.

- RESULTS: A total of 1 Vertebroplasty, 2 ORIF Pelvis fracture, 1 ORIF with PFN for Proximal Femoral Fracture, 1 CRIF for distal radius fracture and 2 ORIF for Tibia Fracture cases are performed to evaluate the benefits of AR surgery. Among the AR surgeries, surgeons wear the smart surgical are lot reduce of eyes of turns to focus on the monitors. This paper shows the potential ability of augmented reality technology for trauma surgery.

Introduction

The research of augmented reality technologies allows surgeons to see the data visualization into diagnostic and treatment procedures to improve workflow easily, safety, and cost and to enhance both surgical training and operation efficiency. However, the awareness of possibilities of augmented reality is generally low. This paper evaluates whether augmented reality can presently improve the results of orthopedic surgical procedures. This technology used an image for each eye to create a virtual image via the transplant glasses for the viewer. In recent years, AR technology has evolved at a rapid pace based on the graphics and computing technologies have evolved. The main purpose of this paper is to see the utilization of AR technology into real-time orthopedic surgeries and with this paper, we will demonstrate the usage of AR technology and how it affects the efficiency of the surgery reducing surgical time and radiation exposure [1].

As the growth of the technology is fast, Virtual Reality (VR), AR and Mixed Reality (MR) technologies have also taken into the medical industry. A lot of research papers on the VR, MR and AR technology discuss the conceptual usage of the technology into the real-time surgeries. New technologies, in particular virtual reality and robotics [2], will have a major impact on health care in the next decade [3]. This describes the application of virtual reality and robotics to surgical training and planning and the execution of procedures in theatre and discusses the near future of this new technology. Virtual reality and robotics are turning out to be a proven technology in recent years but they still have some limitations which restrict the usage of VR into the real cases during the surgical procedure in some of the areas of the surgery. Virtual reality in surgery and, more specifically, in surgical training, faces several challenges in the future. These challenges are building realistic models of the human body, creating interface tools to view, hear, touch, feel, and manipulate these human body models, and integrating virtual reality systems into medical education and treatment [4].

VR technology is to generate a completely virtual environment and while wear the glasses that can fall into the virtual 3D world. The technology represents an emerging technology that can teach surgeons new procedures and can determine their level of competence before they operate on patients. Laparoscopic surgery is an operative technique that requires the surgeon to observe the operation on a video-monitor and requires the acquisition of new skills. The surgical training is an apprenticeship requiring close supervision and 5 to 7 years of training. This paper plan to create two important tasks in a VR simulator and validate their use. The tasks consist of laparoscopic knot-tying and laparoscopic suturing. They proposed a VR in combination with fuzzy logic can educate surgeons and determine when to perform surgical procedures on patients [5]. However, VR technology is not appropriate for real-time orthopedics because the surgeon cannot see through the patient and has the view of the virtual image.

The AR is another technology allows physicians to incorporate data visualization into diagnostic and treatment procedures to improve surgical procedure efficiency, safety and to improve surgical operation and training. This reference paper [6] evaluates whether augmented reality can presently improve the results of surgical procedures. AR comprises special hardware and software, which is used to offer computer-processed imaging data to the surgeon in real-time so that virtual objects from the medical images are combined with computer-generated images [7]. AR technology has recently gained increasing interest not only in-game but also in surgical practice. AR has been applied to a wide spectrum of orthopedic procedures, such as tumor resection [8], fracture fixation [9], arthroscopy [10], and component's alignment in total joint arthroplasty [11]. The use of computed tomography and magnetic resonance images during preoperative planning and intraoperative surgical navigation is vital to the success of the surgery and positive patient outcome. AR application in orthopedic and neurosurgery has the potential to revolutionize and change the way surgeons plan and perform surgical procedures in the future.

A medical industrial device targeting the orthopedic trauma market, Foresee-X [12] helps surgeons by enhancing the synching of images from intra-operative fluoroscopy. This is especially beneficial for pelvic procedures, interlocking nail procedures, etc. The smart surgical glasses which is equipped with an augmented reality solution for visual aid, also has image enhancement functions such as the ability to zoom in and out. The smart surgical glasses can thus increase efficiency by allowing surgeons to concentrate on the operational field instead of monitors. It reduces radiation exposure for surgeons by lowering the necessity of taking X-ray images during operation. It is thus a device that contributes to the safety of the medical staff and patients. Another advantage is that it improves accuracy by tracking the movements of surgical tools such as the puncture needle, trocar, etc.

In the mixed reality technology, a new generation of technology has attracted much attention in recent years [13]. Technology advances have enabled technology to gain increased recognition in medical applications, especially in surgery. Technology will undoubtedly play a significant role in the future, assist surgeons in safely and effectively completing more risky operations. The promise of such technology in the surgical field is huge, as it allows the surgeon (i) to gain access to computer-based solutions in real-time during the procedure while remaining totally sterile, (ii) to gain access to 3-D holograms related to the patient imaging or the surgical technique, and (iii) to remotely interact with colleagues located outside the theatre [14]. The mixed reality is reportedly giving surgeons a superiority rivaling any fictional character, especially the surgeon with the superimposed 3D CT objects on the patient. The latest offering is expected to do big things in the spinal navigation [15]. They develop the AR and MR smart glasses and used to position particular entry points and they can reveal the necessary angles for the screws to be attached for spinal surgery. The MR technology claims that the smart surgical glasses mean that a surgeon will be able to see the exact position of the needle without ever taking their eyes off of the patient. As we can see there are a lot of evidence showing the emergence of VR, MR and AR technologies in recent years for the medical surgeries, so in order to facilitate more evidence, we used AR technology-based smart surgical glasses to perform surgical procedure over various cases to check the effectiveness of the AR technology in the trauma surgeries.

## WORKFLOW FLOWCHART

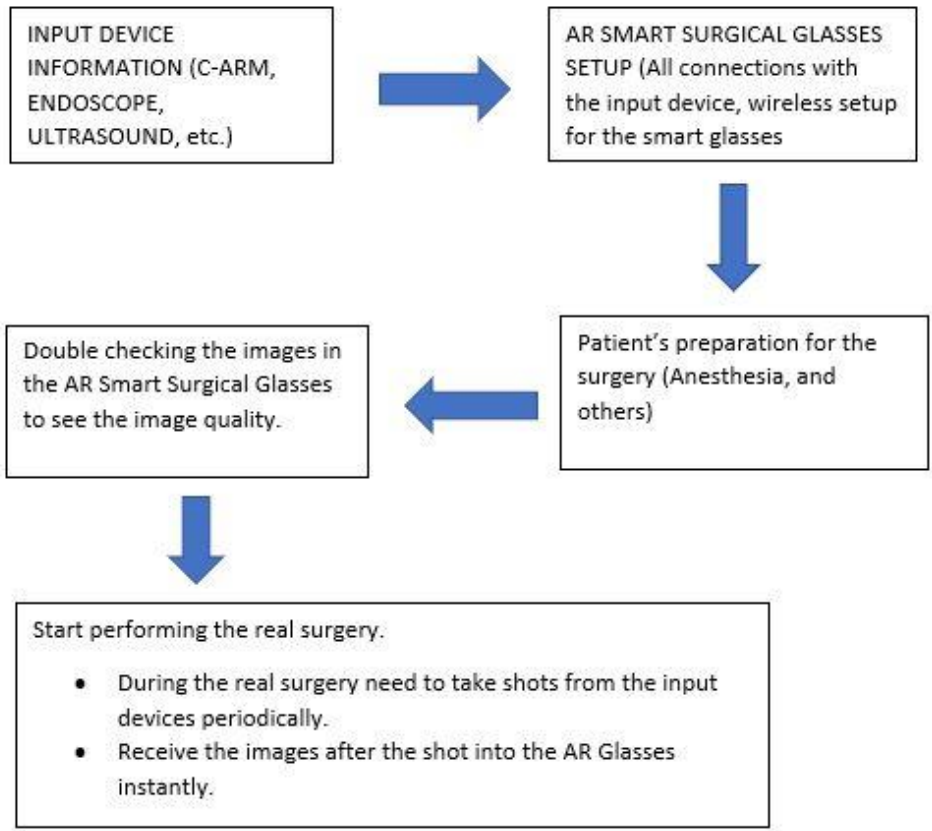

| Using Foresee-X | ORIF for Pelvis Fracture | Vertebroplasty | ORIF with PFN for Proximal Femoral Fracture | ORIF for Tibia Fracture | CRIF for Distal Radius Fracture |
|---|---|---|---|---|---|
| No. Of Cases | 2 | 1 | 1 | 2 | 1 |
| No. of Fluoroscopy | 127 | 90 | 134 | 84/case | 45 |
| Surgical time (mins) | 85/case | 65 | 95 | 82 | 36 |
| Turn of head (Surgeons with glasses) | 40/case | 18 | 52 | 32/case | 6 |
| Turn of head (surgeon's assistant without glasses) | 360/case | 46 | 440 | 72/case | 22 |

The flowchart 1 is the workflow of setup augmented reality for trauma surgery and table 1 shows the number of clinical cases used the smart surgical glasses for trauma cases with the details as shown in Figure 1 to Figure 3.

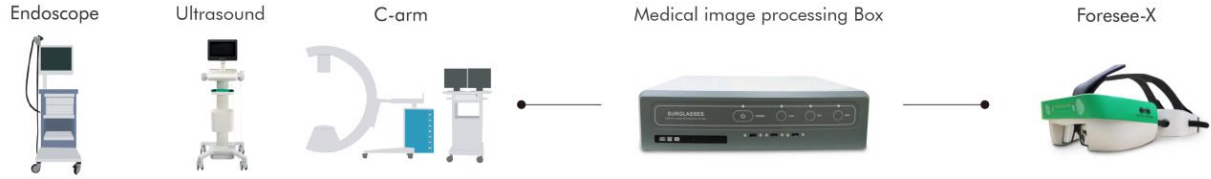

Figure. 1: The system can easily to connect to Endoscopy, Ultrasound and C-arm machines to display the surgical images on smart surgical glasses.

The Workflow and AR System

This paper proposed a workflow to perform AR surgery by using smart surgical glasses. The AR system with a medical image processing box and a AR glasses for only one surgeon used during surgeries. First, set up in the OP room and connect to the C-arm machines (OEC-9900 Elite and Ziehm 8000 in our OP). Second, using HDMI to connect to the image processing box. The AR images are then generated and transmitted to the pair of smart glasses wireless. The displaying resolution of the images is 720p of each glass and the images displaying latency with 30ms under 33Mbs internet speed. The camera sensor is with Full HD resolution. Third, let the patient ready for surgery with anesthesia if needed. Fourth, taking the images with fluoroscopy and see the images by AR glasses see Figure 2. During the consideration, the smart surgical glasses with AR technology was used to perform the surgeries in the real-world surgeries in different hospitals over a variety of Trauma cases, e.g., shoulder, pelvic, endoscope integrated surgeries and ultrasound integrated surgeries.

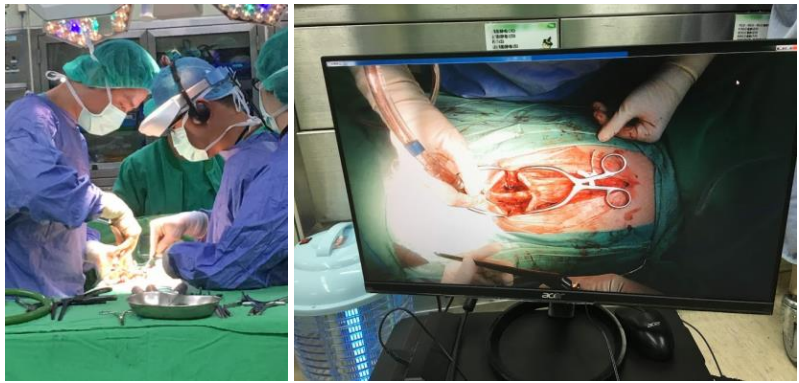

Figure 2: AR Technology smart surgical glasses usage for Real Surgery.

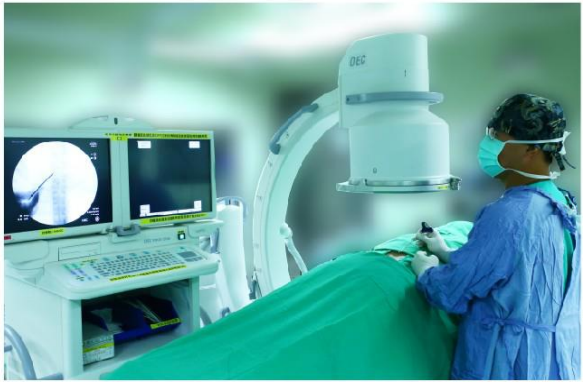 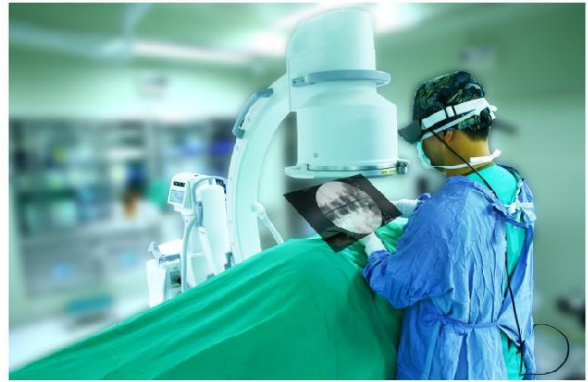

The surgeon needs to focus on the monitors.

Real-time Surgery Image will transfer to the glasses, Doctor will focus on the patients.

Figure 3: Difference in the focus of the surgeon during traditional surgery and surgery with AR smart surgical glasses. The surgeons are not required to focus on eternal monitors too frequently to check the fluoroscopy images. They only need an AR technology smart surgical glasses to see the image pop up in front of the surgical area.

Experimental Results

An augmented reality system helps the surgeons during the surgical procedures with and/or without the processed images obtained from the CT/MRI input with the use of hardware and software. With the help of the Head Mounted Glasses, which we also call smart surgical glasses, the processed image is dynamically transferred to the smart surgical glasses for the surgeons to view with no need to shift focus on the external monitors. Refer to Figure 1. for the understanding of the basic concept of AR technology. The advantages of AR include no external monitors which reduce the frequency of moving head of the surgeon from patient to monitors considerably.

Currently, the interpretation of the AR is under constraint as it depends entirely on the generation of 3D medical images which can be done using the 2D to 3D conversation software available in the market or self-made conversion software. The better the quality of the 2D data, the better is the result for the 3D generation of the images. The 3D generation of the data enables the surgeons for the better understanding of the affected areas, to decide the approach of the surgery to be performed and helps in deciding the entry points for the tools during surgical procedures.

The data to be displayed or seen during the AR surgical procedure depends on the requirement for the surgery and as per the suggestion of the surgeons or the team which they deem required. The requirement of AR into the surgeries are increasing as it facilitates the visualizing of major and critical structures of the human anatomy. An important aspect of AR is its ability to control the data to be displayed or projected on the smart surgical glasses. With the smart surgical glasses, the surgeons can not only see the images from the image processing box into the glasses but also can see the real environment near them. As the technology progresses, the control of the AR smart surgical glasses also enhanced and can be controlled by voice command or hands-free devices which enhances the control of the surgeons on the smart surgical glasses without losing focus on the patients.

Almost all the surgeries, irrespective of the part of the human body, require constant monitoring during the entire surgical procedure especially the inevitable changes in the tissue during the operation. To keep a check, the DICOM images taken during the surgical procedure needs to be transferred instantly to the smart surgical glasses. To observe the situation, the surgeons prefer to take the C-arm scans to keep track of the changes which expose not only the patient but also the surgeons and assisting staff for radiation exposures.

As for the surgical procedures, the complexity of every surgery differs entirely from one another. With the help of AR smart surgical glasses, this problem can be overcome very easily as using AR technology enables the surgeons to have the slightest details of the surgery right in front of them. As technology is progressing into every field, AR technology makes it convenient for the medical and surgical field convenient too. With the AR technology, it reduces the surgical time significantly along with the reduction in radiation exposure for both patient and surgeon and the team. With no need for external monitors, the surgeons can focus more on the operation of the patient. Statistically, the number of times the surgeons move their head during a traditional surgery and the surgery using the AR smart surgical glasses reduces considerably, increasing the efficiency of the surgeons. Taking AR technology into consideration, the scope of the usage of the technology varies into a broader spectrum of the surgical procedures adding to the integration of different apparatuses used in the trauma application like an endoscope, ultrasound [16], etc. With the possibility of using the AR technology over a broader application, AR smart surgical glasses proves to be an excellent assistance for the surgeons.

Fluoroscopic X-ray guidance is a cornerstone for percutaneous orthopedic surgical procedures [8]. However, two-dimensional observations of the three-dimensional anatomy suffer from the effects of projective simplification. Consequently, many X-ray images from various orientations need to be acquired for the surgeon to accurately assess the spatial relations between the patient's anatomy and the surgical tools. In this paper, we present an on-the-fly surgical support system that provides guidance using augmented reality and can be used in quasi-unprepared operating rooms. The proposed workflow with an AR smart surgical glasses system builds upon the medical image processing box to co-calibrate an optical see-through head-mounted display to a C-arm fluoroscopy system. Then, annotations on the 2D X-ray images can be rendered as virtual images or 3D objects providing surgical guidance. We quantitatively evaluate the components of the proposed system, and finally, design a feasibility study on a semi-anthropomorphic phantom. The accuracy of our system was comparable to the traditional image-guided technique while substantially reducing the number of acquired X-ray images as well as procedure time. Our promising results encourage further research on the interaction between virtual and real objects, that we believe will directly benefit the proposed method. Further, we would like to explore the capabilities of us on-the-fly augmented reality support system in a larger study directed towards common orthopedic interventions.

The AR technology smart surgical glasses for ORIF(open reduction and internal fixation) with PFN(proximal femoral nail) for proximal femoral fracture was used during real-time surgery as shown in Figure 4. During the surgery, the C-ARM shots were taken and transferred directly to the smart surgical glasses for the surgeons. As in Figure 4, the last image demonstrates the third person view which can also be used as the training for the resident surgeons who are not inside the operation theater. Ankle surgery too was performed by the surgeons using the AR smart surgical glasses system as shown in Figure 5. Although the AR system has the capability to display the C-ARM images, it depends on the preference of the surgeons sometimes if they want to have the CT data as well during the surgical procedure. For the CRIF(Close Reduction and Internal Fixation) for distal radius fracture, the surgeons used the AR system as a human navigator where they can just tilt their head to find the exact location for the point to be treated as shown in Figure 6. For the vertebroplasty surgery, the C-ARM and AR system was set up and the fluoroscope images were taken during the surgical procedure as shown in Figure 7. As shown in the figure, the view of the surgeon is blocked by the C-ARM and the surgeon cannot focus on the external monitors, thus AR system plays a vital role in the surgical procedure as the images taken during the surgery are transferred to the AR system instantly at the time of shot which lets the surgeons perform seamless surgery. An ORIF for Pelvis fracture was also performed during the testing of the AR Technology Smart Surgical Glasses was done as shown in Figure 8. Lastly, the tibia surgery was performed as shown in Figure 9. As shown in the figure, the tibia locking plate was used during the surgical procedure.

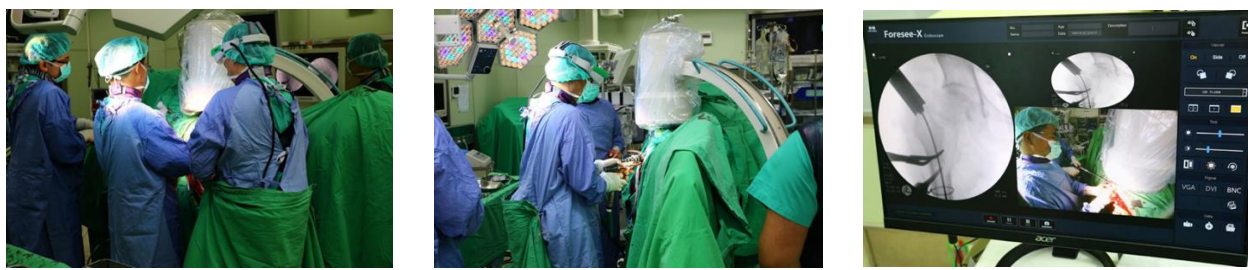

Figure. 4. The AR technology smart surgical glasses connected with the C-ARM for ORIF(open reduction and internal fixation) with PFN(proximal femoral nail) for proximal femoral fracture.

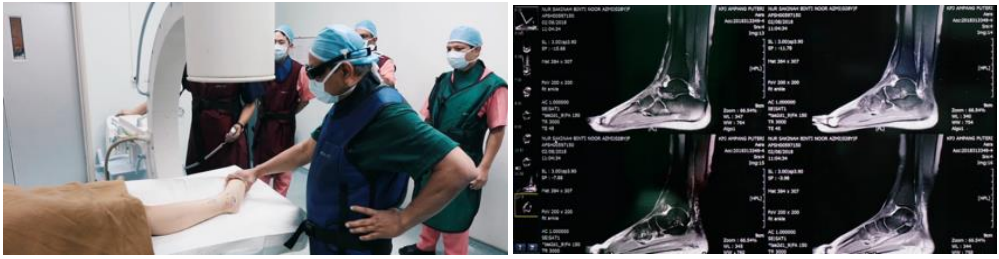

Figure. 5: The AR technology smart surgical glasses connected with the C-ARM for pelvic surgery and also displays CT data as per the surgeon's preference.

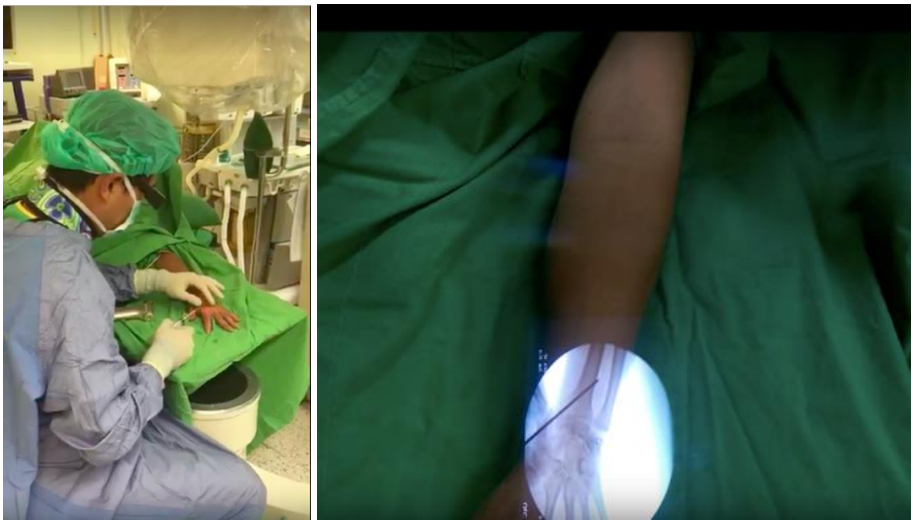

Figure. 6: The AR technology smart surgical glasses connected with the C-ARM for CRIF(Close Reduction and Internal Fixation) for distal radius fracture.

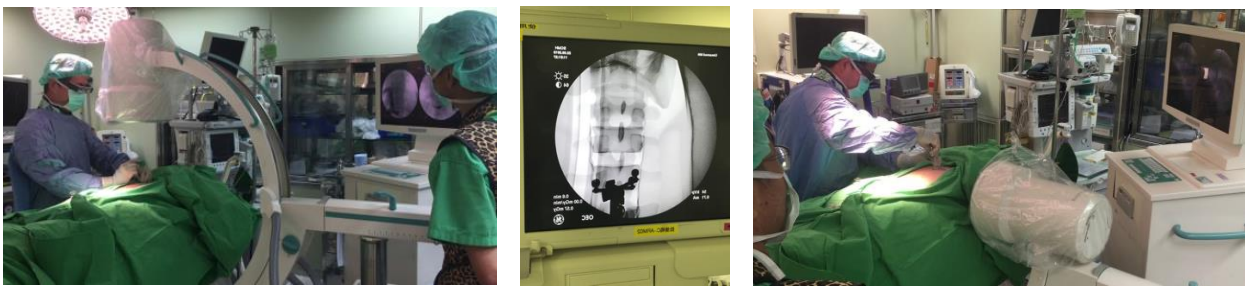

Figure. 7: The AR technology Smart surgical glasses connected with the C-ARM for Vertebroplasty surgery.

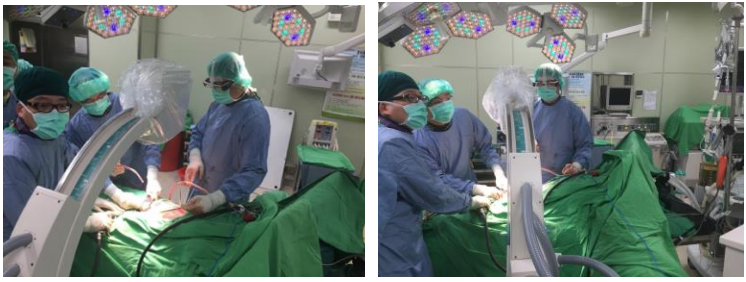

Figure. 8. The AR technology smart surgical glasses connected with the C-ARM for ORIF for Pelvis fracture.

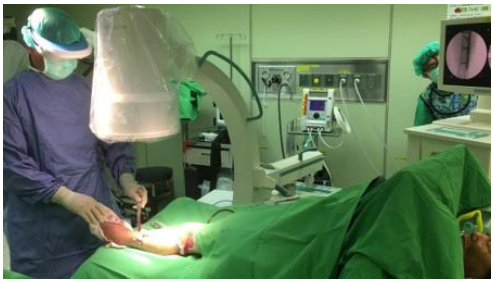

Figure 9: The AR technology smart surgical glasses connected with the C-ARM for ORIF Tibia Fracture

Discussions

After the experimental result, the points that are addressed in this paper according to work flow mentioned at the starting of the paper are followed perfectly. During the whole surgical procedure performed including spine, tibia, hand and pelvic, the AR technology proves to be the future of the surgical procedures. Although AR technology based smart surgical glasses have their advantages, it does have few drawbacks also. The Head Mounted Display is comparatively heavier which makes it a bit inconvenient for long hour surgeries. Other point is the problem coordinating with the assistants available during the surgical procedure. As for now, the image processing box has the capability to connect only one glasses at a time which makes only the surgeon operating the patient has the AR view but the surgeons assisting the surgery still have to rely on the external monitors.

As per the experimental results, AR technology can be used in a wider spectrum of the surgeries and not restricting it to only trauma surgeries. With the capability of the AR technology of the integration with different standard medical devices like ultrasound, endoscope, etc., this technology can provide an immense increase in the efficiency of the surgeries and surgeons. The advantages of the surgery are not restricted to the surgeons only but also to the patients, resident doctors and the surgery assistant personnel. AR technology is very helpful in reducing the radiation exposure of the surgeons, patients and the surgery assistance personnel. Taking into consideration average of all the parameters considered during surgery, the statistics are very different and these parameters differs from the traditional surgeries considerably.  The performance of average collection of parameters from different surgeries performed during the experiments. The average assistant turns his head: 440 times, turning of main surgeon 52 times with C-Arm times: 134 times and the average of AR device used time 95 minutes.

Conclusions

As per the observation from different surgical procedures using the AR technology based smart surgical glasses proves a vital tool in revolutionizing the surgical procedure. The result obtained during the surgery with the use of smart surgical glasses differs significantly from the traditional surgical procedure. The use of smart surgical glasses not only increases the efficiency of the surgeons but also helps the patients to recover faster as the incisions made in the surgical procedure are comparatively smaller than the traditional surgeries. The time required for the surgeries are relatively shorter than the traditional surgeries which let the surgeons cure more patients compared to traditional surgeries which are a major breakthrough for surgeons and also the patients. The accuracy of the surgery is another important factor that can be mentioned as the outcome of the different surgical procedures with a reduction in the surgical procedure timing for up to 30%. Smart surgical glasses can also be used for the educational purpose for the resident doctors where the real-time surgical procedure can be broadcasted outside the surgical theaters or recorded and used for the resident doctors and the aspiring surgeons to study the case studies of different cases from experienced surgeons.

Acknowledgements

This paper was supported by Taiwan Main Orthopaedic Biotechnology Co., Ltd. in 2016-2020.